\documentclass{ws-procs975x65}

\def\be{\begin{equation}}
\def\ee{\end{equation}}

\begin{document}

\title{Recent developments in string model-building and cosmology}
\author{Michele Cicoli}

\address{Dipartimento di Fisica ed Astronomia, Universit\`a di Bologna, \\ via Irnerio 46, 40126 Bologna, Italy \\
$^*$E-mail: michele.cicoli@unibo.it}

\address{INFN, Sezione di Bologna, Italy}

\address{Abdus Salam ICTP, Strada Costiera 11, Trieste 34014, Italy}

\begin{abstract}
In this talk I discuss recent developments in moduli stabilisation, SUSY breaking and chiral D-brane models together with several interesting features of cosmological models built in the framework of type IIB string compactifications. I show that a non-trivial pre-inflationary dynamics can give rise to a power loss at large angular scales for which there have been mounting observational hints from both WMAP and Planck. I then describe different stringy embeddings of inflationary models which yield large or small tensor modes. I finally argue that reheating is generically driven by the decay of the lightest modulus which can produce, together with Standard Model particles, also non-thermal dark matter and light hidden sector degrees of freedom that behave as dark radiation.
\end{abstract}


\bodymatter

\section{Introduction}

During this talk I will focus on type IIB on CY orientifolds with D3/D7-branes and O3/O7-planes since:
\begin{itemize}
\item D-branes provide \emph{non-Abelian} gauge symmetries and \emph{chiral} matter. They can therefore be used to realise MSSM- or GUT-like models via either magnetised D7-branes wrapped around 4-cycles or D3-branes at singularities;

\item Most of the moduli can be fixed with control over moduli space by turning on background fluxes $H_3$, $F_3$ which lead to a small back-reaction on the internal geometry;

\item Type IIB compactifications allow to realise a brane-world scenario where gauge interactions are localised. Thus model-building, being a \emph{local} issue, decouples (at leading order) from moduli stabilisation which is a \emph{global} issue. 
\end{itemize}
The table below shows different local (brane) and global (bulk) issues. The local ones are more model-dependent since they involve the details of particular brane set-ups, while global issues are more model-independent since they are affected by the properties of the bulk of the extra dimensions. Interestingly, some issues like reheating, dark radiation and dark matter are both local and global since they involve the coupling of closed string modes to open string degrees of freedom. 

\begin{center}
{\tablefont
\begin{tabular}{ll}
\toprule
Local (brane) issues & Global (bulk) issues \\\colrule
Gauge group & Moduli stabilisation \\
Chiral spectrum & Cosmological constant \\
Yukawa couplings & Hierarchies \\
Gauge coupling unification & Moduli spectrum \\
Mixing angles & SUSY breaking and soft terms \\
Proton stability & Inflation \\
Reheating & Reheating \\
Dark radiation & Dark radiation \\
Dark matter & Dark matter \\\botrule
\end{tabular}}\label{Tab1}
\end{center}

Recently there has been a lot of progress in trying to combine global with local issues in compact Calabi-Yau models with explicit brane set-up, tadpole cancellation and dS moduli stabilisation compatible with chirality.\cite{Cicoli:2011qg} Let us summarise the main results within the framework of type IIB Large Volume Scenarios which we will discuss in this talk: 
\begin{itemize}
\item Construction of explicit compact Calabi-Yau orientifolds via toric geometry;
\item Presence of an explicit set-up with D3/D7-branes, O3/O7-planes and fluxes (both background and gauge);\cite{Cicoli:2011qg,Cicoli:2012vw,Cicoli:2013mpa,Cicoli:2013zha}
\item Global consistency of the underlying construction due to D3-, D5- and D7-tadpole and Freed-Witten anomaly cancellation;
\item Explicit fixing of dilaton and complex structure moduli by reducing the effective number of moduli due to symmetries in the moduli space identified using the Greene-Plesser construction;\cite{Cicoli:2013cha}
\item Stabilisation of the K\"ahler moduli in a way compatible with chirality within regime of validity of the effective field theory;
\item Two different realisations of the visible sector:
\begin{enumerate}
\item D7-branes in geometric regime can lead to chiral $SU(5)$- or MSSM-like models,\cite{Cicoli:2011qg}
\item Fractional D3-branes and flavour D7-branes at del Pezzo singularities can accommodate $SU(3)^3$, Pati-Salam or MSSM-like models,\cite{Cicoli:2012vw,Cicoli:2013mpa,Cicoli:2013zha,Cicoli:2013cha}
\end{enumerate}
\item dS vacua without anti-branes via two fully supersymmetric methods:
\begin{enumerate}
\item Non-zero F-terms of charged hidden matter fields induced by D-term stabilisation,\cite{Cicoli:2012vw,Cicoli:2013mpa,Cicoli:2013cha}
\item Non-perturbative effects at singularities,\cite{Cicoli:2012fh}
\end{enumerate}
\item Spontaneous SUSY breaking by F-terms of K\"ahler moduli;\cite{Aparicio:2014wxa}
\item TeV-scale soft-terms via gravity mediation;
\item The K\"ahler moduli are promising inflaton candidates since:\cite{Burgess:2013sla}
\begin{enumerate} 
\item The $\eta$-problem can be solved by the extended no-scale structure of the K\"ahler potential,
\item Explicit stringy realisations of $\alpha$-attractors with inflationary potentials of the schematic and generic form $V\simeq V_0\left(1-k\,e^{-k\phi}\right)$,
\item Possible power loss at large angular scales due to a non slow-roll pre-inflationary evolution,\cite{Cicoli:2013oba, Cicoli:2014bja} 
\end{enumerate}
\item Reheating is driven by the decay of the lightest modulus;\cite{Allahverdi:2013noa}
\item Generic production of non-thermal neutralino dark matter,\cite{Allahverdi:2014ppa} and axionic dark radiation;\cite{Cicoli:2012aq}
\item Prediction of the existence of a cosmic axion background with $\mathcal{O}$($200$ eV) energies;
\item Possible explanation of the observed soft X-ray excess,\cite{Conlon:2013txa} and $3.5$ keV line from galaxy clusters due to axion-photon conversion in the cluster magnetic field.\cite{Cicoli:2014bfa}
\end{itemize}

\section{Moduli stabilisation and SUSY breaking}

In this section we give a very brief review of the main aspects of dS closed string moduli stabilisation and SUSY breaking. 

\subsection{Tree-level stabilisation}

The 4D type IIB tree-level K\"ahler potential $K$ and superpotential $W$ read:
\be
K_{\rm tree}= -2\ln\mathcal{V}(T_i+\bar{T}_i)-\ln(S+\bar{S})-\ln\left(i\int_{\rm CY}\Omega(U)\wedge \bar{\Omega}\right)
\qquad W_{\rm tree} = \int_{\rm CY} G_3\wedge \Omega(U) \nonumber
\ee
leading to a scalar potential $V =e^K\left[K^{i\bar{j}}D_i W D_{\bar{j}}\bar{W}-3|W|^2\right]$ of the form: 
\be
V = e^K \sum_{S,U}K^{\alpha\bar{\beta}}D_\alpha W D_{\bar{\beta}}\bar{W}+e^K\left[\sum_T K^{i\bar{j}} K_i K_{\bar{j}}-3\right]|W|^2 \geq 0 \nonumber
\ee
due to the no-scale cancellation $\sum_T K^{i\bar{j}} K_i K_{\bar{j}}=3$. The dilaton $S$ and the complex structure moduli $U$ can be fixed supersymmetrically at $D_S W = D_U W=0$ setting $W_0 =\langle W_{\rm tree}\rangle$. These are $n=2 h^{1,2}+2$ real non-linear equations in $n$ unknowns with $2n$ parameters, the flux quanta, whose values are constrained by D3 tadpole cancellation. There is therefore enough freedom to find solutions whose number can be estimated as follows. If each flux quanta can take for example $10$ different values we have:
\be
N_{\rm sol}\sim 10^{2n}= 10^{4(h^{1,2}+1)} \sim 10^{400}\quad\text{for}\quad h^{1,2}\sim \mathcal{O}(100) \nonumber
\ee
This leads to the flux landscape. At this level of approximation the vacuum is Minkowski and SUSY is broken since $F^T \neq 0$. However the $T$-moduli are still flat. The gravitino mass turns out to be:
\be
m_{3/2} = e^{K/2} |W| \simeq \frac{W_0}{\mathcal{V}}\,M_p
\ee
Notice that natural values of the underlying parameters lead to $W_0 \sim \mathcal{O}(1)$ while $W_0\ll 1$ requires some fine-tuning of the flux quanta. 

\subsection{K\"ahler moduli stabilisation}

$K$ and $W$ get corrected beyond tree-level as follows: 
\be
W=W_{\rm tree} + W_{\rm np} \qquad K \simeq K_{\rm tree} + K_{\rm p} \quad\text{with}\quad K_{\rm p} = K_{\alpha'}+K_{g_s} \nonumber
\ee
where:
\be
W_{\rm np}\sim e^{-\tau}\ll K_{\rm p}\sim K_{\alpha'}\sim \frac{1}{\mathcal{V}}\sim \frac{1}{\tau^{3/2}}\quad\text{for}\quad \tau\gg 1 \nonumber
\ee
It follows that $V=V_{\rm tree}+V_{\rm p}+V_{\rm np}$ where $V_{\rm tree}=0$ due to the no-scale cancellation. On the other hand the scaling behavior of $V_{\rm p}$ and $V_{\rm np}$ is:
\be
V_{\rm p} \sim e^K W_0^2 K_{\rm p} \qquad V_{\rm np}\sim e^K (W_0 W_{\rm np} + W_{\rm np}^2)  \nonumber
\ee
In the natural case with $W_0\sim \mathcal{O}(1)$ we have:
\be
\frac{V_{\rm p}}{V_{\rm np}} \sim \frac{K_{\rm p}}{W_{\rm np}} \sim \frac{e^\tau}{\tau^{3/2}}\gg 1 \quad\Rightarrow\quad V_{\rm p}\gg V_{\rm np} \nonumber
\ee
Thus non-perturbative effects can be neglected and moduli stabilisation has to take place at perturbative level. However this can be done only via tuning since $V_{g_s}\ll V_{\alpha'}$ due to the extended no-scale structure.\cite{Cicoli:2007xp} On the other hand, if we tune $W_0\sim \mathcal{O}(W_{\rm np})$ we have:
\be
\frac{V_{\rm p}}{V_{\rm np}} \sim K_{\rm p} \sim \frac{1}{\tau^{3/2}}\ll 1 \quad\Rightarrow\quad V_{\rm p}\ll V_{\rm np} \nonumber
\ee
In this case we have therefore to perform a pure non-perturbative KKLT-like stabilisation. However a full non-perturbative fixing has some shortcomings: 
\begin{itemize}
\item $W_0$ has to be tuned;
\item $W_{\rm np}$ gets definitely generated for \emph{rigid} cycles while non-perturbative effects for non-rigid cycles are not guaranteed;
\item There is a tension between moduli stabilisation and chirality which can be schematically summarised as follows:\cite{Blumenhagen:2007sm}
\begin{enumerate}
\item If the visible sector wraps the 4-cycle $\tau$ with gauge flux $F$, $\tau$ gets a $U(1)$ charge
\item If $\tau$ is also wrapped by an E3-instanton, the contribution $W_{\rm np}\sim e^{-\tau}$ would not be gauge invariant
\item Chiral intersections between the E3 and the visible sector make $W_{\rm np}\sim \prod_i \phi_i \,e^{-\tau}$ gauge invariant
\item In order to preserve visible sector gauge symmetries at high energies we need $\langle\phi_i\rangle=0$ $\forall i$. This in turn gives $W_{\rm np}=0$, implying that the 4-cycle supporting the visible sector cannot be fixed by non-perturbative effects
\end{enumerate}
\item There is a tension also between moduli stabilisation and Freed-Witten anomaly cancellation which we briefly summarise as:
\begin{enumerate}
\item In the simplest case in order to generate a non-zero $W_{\rm np}$, an E3 has to wrap a transversally invariant cycle with $\mathcal{F}=F-B=0$
\item In the case of non-spin cycles, Freed-Witten anomaly cancellation induces $F_{\rm FW}=\frac 12 \hat{D}\neq 0$
\item One can therefore choose the $B$-field to cancel $F_{\rm FW}$ as $B = F_{\rm FW}$. However then it becomes hard to cancel $F-B$ simultaneously for two or more intersecting cycles 
\end{enumerate}
\item Standard KKLT-like stabilisation procedures lead to AdS vacua where the dS uplifting is performed with anti D3-branes. The consistency of these constructions is presently under debate.
\end{itemize}

\subsection{Large Volume Scenario}

A very elegant way-out to the previous problems can be found if $h^{1,1} \geq 2$. In fact in this case one can fix the moduli without tuning since for 
$W_0 \sim \mathcal{O}(1)$ we have:
\be
\frac{V_{\rm p}}{V_{\rm np}} \sim \frac{K_{\rm p}}{W_{\rm np}} \sim \frac{e^{\tau_s}}{{\tau_b}^{3/2}}\sim \mathcal{O}(1) \quad\text{if}\quad 1\ll\tau_s\ll\tau_b \nonumber
\ee
This is obtained dynamically if $\tau_s$ is a diagonal blow-up mode and the internal volume is of the form $\mathcal{V}=\tau_b^{3/2}-\tau_s^{3/2}\simeq \tau_b^{3/2}$. This is a very promising situation since:
\begin{enumerate}
\item $\tau_s$ is a local effect, and so it naturally reproduces the required hierarchy $\tau_s \ll  \tau_b$
\item $\tau_s$ is a rigid cycle, and so $W_{\rm np}$ gets easily generated.
\end{enumerate}
This leads to the Large Volume Scenario where the K\"ahler potential and the superpotential look like ($\xi$ is an $\mathcal{O}(1)$ topological quantity controlling $\alpha'$ effects):
\be
K=-2\ln\mathcal{V}-\frac{\xi}{g_s^{3/2}\mathcal{V}}\qquad W=W_0+A_s\,e^{-a_s T_s} \nonumber
\ee
The scalar potential after axion minimisation becomes (the $\lambda$'s are $\mathcal{O}(1)$ constants):
\be
V= \lambda_1 \sqrt{\tau_s}\frac{e^{-2a_s\tau_s}}{\mathcal{V}}- \lambda_2 \tau_s W_0\frac{e^{-a_s\tau_s}}{\mathcal{V}^2}+\lambda_3\frac{W_0^2}{g_s^{3/2}\mathcal{V}^3} 
\label{VLVS}
\ee
admitting an AdS minimum at: 
\be
\tau_s\sim g_s^{-1}\sim\mathcal{O}(10)\quad\text{for}\quad g_s\simeq 0.1 \qquad \mathcal{V}\simeq W_0\,e^{a_s\tau_s}\sim e^{1/g_s}\gg 1 \nonumber
\ee
The presence of an exponentially large internal volume allows to trust the approximations and to generate hierarchies naturally. For example, the gravitino mass becomes exponentially suppressed with respect to the Planck scale:
\be
m_{3/2}\simeq \frac{W_0}{\mathcal{V}} M_p \simeq M_p\,e^{-1/g_s} \ll M_p \nonumber
\ee
Hence this scenario can yield low-energy SUSY naturally. Moreover SUSY is spontaneously broken since at the minimum: 
\be
F^{T_b}\sim \frac{M_p^2}{\mathcal{V}^{1/3}}\neq 0\qquad\qquad F^{T_s}\sim \frac{M_p^2}{\mathcal{V}}\neq 0
\nonumber
\ee
The Large Volume Scenario also does not feature any conflict between non-perturbative effects and chirality and Freed-Witten anomaly cancellation since: 
\begin{enumerate}
\item $\tau_s$ does not intersect other cycles, and so there are no chiral intersections between the visible sector and instantons on $\tau_s$, resulting in a non-zero $W_{\rm np}$. The visible sector cycle has instead to be fixed by either D-terms or $g_s$ effects; 
\item In order to fix the moduli one needs only non-perturbative effects for diagonal blow-up modes. In this case one can choose the $B$-field to cancel all Freed-Witten fluxes since non-perturbative effects are supported on non-intersecting cycles.
\end{enumerate}

In the Large Volume Scenario the minimal number of K\"ahler moduli to be realistic is $h^{1,1}\geq 4$. In fact, there are two possible ways to realise the visible sector:
\begin{itemize}
\item In models with D7-branes in the geometric regime, the internal volume looks schematically as:\cite{Cicoli:2011qg}
\be
\mathcal{V}=\tau_b^{3/2}-\tau_s^{3/2}-\left(\tau_{{\rm vs}_1}+\tau_{{\rm vs}_2}\right)^{3/2}
\nonumber
\ee
At leading order D-terms fix $\tau_{{\rm vs}_1}\sim \tau_{{\rm vs}_1}$ leaving a flat direction which we call $\tau_{\rm vs}$. Subdominant non-perturbative and $\alpha'$ effects fix $\tau_b$ and $\tau_s$ at $\tau_b^{3/2}\sim e^{\tau_s}$ and $\tau_s\sim g_s^{-1}$. Finally $g_s$ effects stabilise $\tau_{\rm vs}$.

\item In models with D3-branes at singularities, the Calabi-Yau volume should take instead the form:\cite{Cicoli:2012vw,Cicoli:2013mpa,Cicoli:2013zha}
\be
\mathcal{V}=\tau_b^{3/2}-\tau_s^{3/2}-\tau_{{\rm vs}_1}^{3/2}-\tau_{{\rm vs}_2}^{3/2}
\nonumber
\ee
The two 4-cycles $\tau_{{\rm vs}_1}$ and $\tau_{{\rm vs}_2}$ are exchanged by the orientifold involution in order to obtain unitary groups for the visible sector. The shrinking of $\tau_{{\rm vs}_1}$ and $\tau_{{\rm vs}_2}$ to zero size is induced by D-term fixing.  At subleading order non-perturbative and $\alpha'$ effects fix $\tau_b$ and $\tau_s$ at $\tau_b^{3/2}\sim e^{\tau_s}$ and $\tau_s\sim g_s^{-1}$. 
\end{itemize}
Let us now analyse separately the phenomenological implications of these two different realisations of the visible sector.

\subsection{Unsequestered models}

When the visible sector is built with D7-branes in geometric regime, the F-term of $\tau_{\rm vs}$ is non-zero: $F^{\rm vs}\sim m_{3/2} M_p\neq 0$. The soft-terms and the mass of the volume mode scale as:          
\be
m_{\mathcal{V}}\sim \frac{M_p}{\mathcal{V}^{3/2}} \ll M_{\rm soft}\sim m_{3/2} \sim \frac{M_p}{\mathcal{V}}
\nonumber
\ee
One can therefore set either $M_{\rm soft}\sim\mathcal{O}(1)$ TeV to solve the hierarchy problem or $m_{\mathcal{V}} > \mathcal{O}(50)$ TeV to avoid any cosmological moduli problem. These two different choices require two different values of the internal volume which in turn set all the other relevant energy scales. The table below shows the values of all these energy scales for $\mathcal{V}\sim 10^{14}$ and $\mathcal{V}\sim 10^4$ respectively.

\begin{center}
{\tablefont
\begin{tabular}{ll}
\toprule
Energy scales for $\mathcal{V}\sim 10^{14}$ & Energy scales for $\mathcal{V}\sim 10^4$ \\\colrule
$M_p\sim 10^{18}$ GeV & $M_p\sim 10^{18}$ GeV \\
$M_s\sim m_{\tau_{{\rm vs}_1}}\sim m_{a_{{\rm vs}_1}}\sim M_p\mathcal{V}^{-1/2}\sim 10^{11}$ GeV & $M_s\sim m_{\tau_{{\rm vs}_1}}\sim m_{a_{{\rm vs}_1}}\sim 10^{16}$ GeV \\
$M_{\rm KK}\sim M_p\mathcal{V}^{-2/3}\sim 10^8$ GeV & $M_{\rm KK}\sim 10^{15}$ GeV \\
$m_{\tau_s}\sim m_{a_s}\sim M_p \mathcal{V}^{-1}\ln\mathcal{V}\sim 100$ TeV & $m_{\tau_s}\sim m_{a_s}\sim 5\cdot 10^{14}$ GeV \\
$m_{3/2}\sim M_p \mathcal{V}^{-1}\sim 10$ TeV & $m_{3/2}\sim 10^{14}$ GeV \\
$M_{\rm soft}\sim m_{\tau_{{\rm vs}_2}}\sim M_p \mathcal{V}^{-1}(\ln\mathcal{V})^{-1}\sim 1$ TeV & $M_{\rm soft}\sim m_{\tau_{{\rm vs}_2}}\sim 10^{13}$ GeV \\
$m_{\tau_b}\sim M_p \mathcal{V}^{-3/2}\sim 1$ MeV & $m_{\tau_b}\sim 10^{12}$ GeV \\
$m_{a_{{\rm vs}_2}}\sim \Lambda_{\rm QCD}^2 f_{a_{{\rm vs}_2}}^{-1}\sim 1$ meV for $f_{a_{{\rm vs}_2}}\sim M_s$ & $m_{a_{{\rm vs}_2}}\sim 1$ neV for $f_{a_{{\rm vs}_2}}\sim M_s$ \\
$m_{a_b}\sim M_p \,e^{-\mathcal{V}^{2/3}}\sim 0$ & $m_{a_b}\sim 0$ \\\botrule
\end{tabular}}\label{Tab2}
\end{center}

\subsection{Sequestered models}

When the visible sector lives on D3-branes at singularities, the F-term of $\tau_{\rm vs}$ vanishes: $F^{\rm vs} \propto \xi_{\rm FI} \propto \tau_{\rm vs} \to 0$. This cancellation induces a sequestering of the visible sector from the sources of SUSY breaking which are the F-terms of the bulk moduli. Thus gaugino masses turn out to be suppressed with respect to $m_{3/2}$:\cite{Aparicio:2014wxa}          
\be
M_{1/2}\sim\frac{M_p}{\mathcal{V}^2}\ll m_{3/2}\sim \frac{M_p}{\mathcal{V}}
\nonumber
\ee
Depending on the exact moduli-dependence of the matter K\"ahler metric and the mechanism responsible to achieve a dS minimum, scalar masses instead scale as:
\be
m_0\sim \frac{M_p}{\mathcal{V}^{3/2}}\sim m_{\mathcal{V}}\qquad\text{or}\qquad m_0 \sim \frac{M_p}{\mathcal{V}^2} \sim M_{1/2}
\nonumber
\ee
Setting $\mathcal{V}\sim 10^7$, one can obtain $M_{1/2}\sim\mathcal{O}(1)$ TeV. All the other main energy scales are listed in the table below.

\begin{center}
{\tablefont
\begin{tabular}{l}
\toprule
Energy scales \\\colrule
$M_p\sim 10^{18}$ GeV \\
$M_{\rm GUT}\sim M_s \mathcal{V}^{1/6}\sim 10^{16}$ GeV
$M_s\sim m_{\tau_{{\rm vs}_1}}\sim m_{a_{{\rm vs}_1}}\sim m_{\tau_{{\rm vs}_2}}\sim m_{a_{{\rm vs}_2}}\sim 10^{15}$ GeV \\
$M_{\rm KK}\sim 10^{14}$ GeV \\
$m_{\tau_s}\sim m_{a_s}\sim 10^{12}$ GeV \\
$m_{3/2}\sim 10^{11}$ GeV \\
$m_{\tau_b}\sim 10^7$ GeV \\
$M_{1/2}\sim 1$ TeV \\
$m_{a_{\rm open}}\sim 1$ meV for $f_{a_{\rm open}}\sim M_s\sqrt{\tau_{\rm vs}}\ll M_s$ \\
$m_{a_b}\sim 0$ \\\botrule
\end{tabular}}\label{Tab3}
\end{center}
Scalar masses can be either $m_0\sim M_{1/2}\sim 1$ TeV as in standard MSSM-like models, or $m_0\sim m_{\tau_b}\sim 10^7$ GeV as in split SUSY-like scenarios. This scenario can therefore allow for: low-energy SUSY, a promising framework to embed standard GUT theories, a right inflationary scale, no cosmological moduli problem for $\tau_b$, a viable QCD axion from open string modes, reheating driven by the decay of $\tau_b$ with $T_{\rm rh} \sim 1-10$ GeV, non-thermal dark matter and axionic dark radiation produced from the decay of $\tau_b$.

\subsection{dS from hidden F-terms}

Let us now briefly present a general mechanism which can lead to dS vacua. In globally consistent models $\tau_b$ is wrapped by a hidden stack of D7-branes because of D7-tadpole cancellation. Moreover Freed-Witten anomaly cancellation induces a non-zero gauge flux on $\tau_b$.\cite{Cicoli:2012vw,Cicoli:2013mpa,Cicoli:2013zha} This modulus therefore acquires a $U(1)$-charge and appears in the Fayet-Iliopoulos term of the D-term potential:
\be
V_D^{\rm bulk}=\frac{1}{\tau_b}\left(\sum_i q_{D7i}|\phi_i|^2-\xi_{D7}\right)^2\quad\text{with}\quad\xi_{D7}=\frac{3}{(2\mathcal{V})^{2/3}}
\nonumber
\ee
The total scalar potential reads:
\be
V_{\rm tot} = V_D^{\rm bulk} + V_F = \frac{1}{\tau_b}\left(q_{D7}|\phi_{\rm dS}|^2-\xi_{D7}\right)^2+m_{3/2}^2 |\phi_{\rm dS}|^2 + V_{\mathcal{O}(\mathcal{V}^{-3})}
\nonumber
\ee
where is $V_{\mathcal{O}(\mathcal{V}^{-3})}$ the moduli potential (\ref{VLVS}). The minimum for $\phi_{\rm dS}$ lies at
\be
q_{D7}|\phi_{\rm dS}|^2 = \xi_{D7}-\frac{m_{3/2}^2\tau_b}{2 q_{D7}}
\nonumber
\ee
Substituting this result in $V_{\rm tot}$ we obtain:
\be
V_{\rm tot} = V_{D,0}^{\rm bulk} + V_F = \frac{m_{3/2}^4 \tau_b}{4 q_{D7}^2} + m_{3/2}^2 \frac{\xi_{D7}}{q_{D7}} + V_{\mathcal{O}(\mathcal{V}^{-3})}
\label{f}
\ee
The first term on the RHS of (\ref{f}) is negligible since it scales as $\mathcal{V}^{-10/3}$ while the second term on the RHS behaves as $\mathcal{V}^{-8/3}$ and can play the r\^ole of an uplifting term. Minimising with respect to $\tau_s$ and $\mathcal{V}$ we obtain
\be
\langle V_{\rm tot}\rangle = \frac{3 W_0^2}{4 a_s^{3/2}\mathcal{V}^3}\left[\delta \mathcal{V}^{1/3} - \sqrt{\ln\left(\frac{\mathcal{V}}{W_0}\right)}\right]
\quad\text{with}\quad \delta \simeq 0.01 \left(\frac{a_s^{3/2}}{q_{D7}}\right)
\nonumber
\ee
Clearly $W_0$ can be tuned to get $\langle V_{\rm tot}\rangle = 0$. In particular, $W_0 \sim \mathcal{O}(1)$ gives rise to solutions around $\mathcal{V}\sim 10^6$-$10^7$ which are the values needed to get TeV-scale SUSY.\cite{Cicoli:2012vw,Cicoli:2013mpa,Cicoli:2013cha}

This uplifting mechanism has an interesting higher dimensional understanding in terms of T-branes.\cite{Cicoli:2015ylx} In fact, the effective field theory has to be expanded around the correct background. For a hidden D7-stack this is parameterised by an adjoint complex scalar $\Phi$. The non-zero gauge flux breaks $SO(8)$ to $U(4)$ (focusing on the case of 4 D7s on top of an O7), and so $\Phi$ decomposes as $\textbf{28}\to \textbf{16}_0 \oplus\textbf{6}_{+2} \oplus \textbf{6}_{-2}$. A deformation of $\Phi$ can be written as
\be
\delta\Phi = \left( \begin{array}{cc}
\phi_{\textbf{16}_0} & \phi_{\textbf{6}_{+2}}  \\
\phi_{\textbf{6}_{-2}} & -\phi^T_{\textbf{16}_0} \end{array} \right)
\nonumber
\ee
The 8D BPS equation of motion for a hidden D7-brane is $J\wedge \mathcal{F}_{D7} + \left[\Phi,\bar{\Phi}\right] d {\rm vol}_4 = 0$, 
implying that if $J\wedge \mathcal{F}_{D7} \neq 0$ for $\mathcal{F}_{D7} \neq 0$, $\left[\Phi,\bar{\Phi}\right] \neq 0$. Thus $\Phi$ cannot be in the Cartan and has to take the simple form:
\be
\langle\Phi\rangle = \left( \begin{array}{cc}
0 & \phi_{\textbf{6}_{+2}}  \\
0 & 0 \end{array} \right)
\nonumber
\ee
This is a T-brane background. The gauge group is broken to $SO(4)$. Given that there is no $U(1)$ left, one should not see any D-term contribution if the effective field theory is expanded around the correct background. However by expanding the brane action around this vacuum expectation value in the presence of background fluxes (soft SUSY-breaking scalar masses) one finds the same uplifting term in (\ref{f}).\cite{Cicoli:2015ylx} 

\section{Inflation}

\subsection{Slow-roll inflation}

The emerging picture from COBE, WMAP, Planck and BICEP is a striking simplicity since:
\begin{enumerate}
\item The scalar fluctuations are Gaussian;
\item The spectral index is almost scale-invariant: $n_s\simeq  0.9655 \pm 0.0062$ at 68\% CL;\cite{Ade:2015xua}
\item There is no evidence for tensor modes: $r < 0.11$ at 95\% CL.\cite{Ade:2015xua}
\end{enumerate}
This picture can be elegantly described by an early epoch of accelerated expansion driven by a scalar field. The most popular scenario is slow-roll inflation which is realised when:
\be
\epsilon\equiv \frac{M_p^2}{2}\left(\frac{V'}{V}\right)^2\ll 1\qquad\qquad \eta\equiv M_p^2\frac{V''}{V}\simeq \left(\frac{m_{\rm inf}}{H_{\rm inf}}\right)^2\ll 1
\nonumber
\ee
During this inflationary epoch one has $V\simeq 3 H_{\rm inf}^2 M_p^2$. The duration of inflation to solve the flatness and horizon problems has to be:
\be
N_e=\frac{1}{M_p}\int^{\phi_{\rm in}}_{\phi_{\rm end}} \frac{1}{\sqrt{2\epsilon}} d\phi\gtrsim 60
\nonumber
\ee
Quantum fluctuations of the inflaton generate the density perturbations whose spectrum looks like:
\be
P_S(k)\simeq A_S^2 k^{n_s-1}\qquad A_S \simeq \frac{H_{\rm inf}}{2\pi\sqrt{2}M_p\sqrt{\epsilon}} \simeq 5\cdot 10^{-5}\qquad n_s-1= 2\eta-6\epsilon\simeq -0.04
\nonumber
\ee
Another important inflationary observable is the tensor-to-scalar ratio $r\equiv A_T^2/A_S^2=16 \epsilon<0.11$. This upper bound can be translated also into 
$H_{\rm inf}< 10^{14}$ GeV and $M_{\rm inf}= V^{1/4}<2\cdot 10^{16}\,{\rm GeV}\simeq M_{\rm GUT}$.\cite{Ade:2015xua}

\subsection{String inflation}

Given that recent Planck data can very well be explained by a simple slow-roll inflationary model with a canonically normalised inflaton field, why should one try to embed inflation in a complicated theory as string theory? Because inflation is UV-sensitive, and so one has to embed it in a complete theory of quantum gravity as string theory in order to trust any inflationary model building. 

The UV-sensitivity of inflation is related to the necessity to obtain abnormally flat potentials. This is the so-called $\eta$-problem which is very similar to the hierarchy problem for the Higgs which asks why $m_H \ll M_p$. Similarly for the inflaton one could ask why $m_{\rm inf}\ll H_{\rm inf}$ if there are no symmetries protecting the inflaton potential and controlling Planck-suppressed operators of the form: 
\be
\Delta V\simeq \lambda V \frac{\varphi^2}{M_p^2}\quad\Rightarrow\quad \Delta m_{\rm inf}^2 \sim \lambda\frac{V}{M_p^2}\sim \lambda H_{\rm inf}^2
\quad\Rightarrow\quad \Delta \eta \simeq 1\quad\text{for}\quad \lambda\sim\mathcal{O}(1)
\nonumber
\ee
Moreover observable gravity waves require a trans-Planckian field motion of the inflaton due to the famous Lyth bound: $\frac{\Delta\varphi}{M_p}\simeq\sqrt{\frac{r}{0.001}}$ which implies $\Delta\varphi> M_p$ for $r>0.001$. Thus in this case the situation gets even worse since without a symmetry with a clear UV origin one cannot trust the low-energy expansion:
\be
V(\varphi)=V_0 + \frac{m^2}{2}\varphi^2 + \varphi^4\sum_{i=0}^\infty \lambda_i\left(\frac{\varphi}{M_p}\right)^i
\nonumber
\ee

String inflationary models mainly divide into two classes:\cite{Baumann:2014nda}
\begin{enumerate}
\item Open string inflation: the inflaton is generically a brane position modulus. There is no symmetry solving the $\eta$-problem, and so all these models involve some degree of fine-tuning. There is also an upper bound on the inflaton range from the size of the extra dimensions which leads to the prediction of undetectable tensor modes.

\item Closed string inflation: the inflaton is in general an axion or a volume modulus. There are approximate symmetries solving the $\eta$-problem and models with detectable tensor modes.
\end{enumerate}
From the previous discussion, we have learned that, in order to trust inflation, the inflaton should be a pseudo Nambu-Goldstone boson with a flat potential (over trans-Planckian distances for large $r$).\cite{Burgess:2014tja} Moreover, all the other fields should be decoupled from the inflationary dynamics (for example by making them heavy). The symmetries that can be used can be: \emph{Abelian} yielding single field inflation, or \emph{non-Abelian} leading to multi-field inflationary models, which are however disfavoured by the non-observation of non-Gaussianities. In turn Abelian symmetries can be either a compact $U(1)$ in the case of axion inflation, or a non-compact rescaling in the case of inflation driven by volume moduli.

\subsection*{Compact Abelian pseudo NG bosons}

In the case of compact Abelian pseudo Nambu-Goldstone bosons, the inflaton is an axion and the symmetry is a $U(1)$:
\be
\Phi\to e^{i\alpha}\,\Phi\qquad \Phi=\rho\,e^{i\theta}\qquad \theta\to\theta+\alpha
\nonumber
\ee
Trading $\theta$ for $\varphi$ via the canonical normalisation $\theta = \varphi/f$, the periodic shift symmetry becomes $\varphi\to\varphi+\alpha f$. This is broken by effects of the form $V_0\, e^{\pm i \varphi/f}$ which give rise to the following inflaton potential:
\be
V=V_0\left[1-\cos\left(\frac{\varphi}{f}\right)\right]\quad\Rightarrow\quad \epsilon, \eta \propto \left(\frac{M_p}{f}\right)^2\ll 1\quad\Leftrightarrow \qquad f>M_p
\nonumber
\ee
It is however very hard to get a trans-Planckian $f$ in a low-energy effective theory which is fully under control. Nevertheless there are complicated models with an effective trans-Planckian $f$ which give large tensor modes of order $r>0.01$.\cite{Pajer:2013fsa} Some technical control issues in these models are:	\cite{Cicoli:2014sva}
\begin{enumerate}
\item Renomalisation of $M_p$ due to $N$ light species running in loops: $\delta M_p^2\sim \frac{N}{16\pi^2} M_p^2$
\item Corrections to the effective field theory
\item Decoupling of all fields orthogonal to the inflationary direction by making them heavier than the inflaton
\end{enumerate}

\subsection*{Non-compact Abelian pseudo NG bosons}

The most common non-compact Abelian pseudo Nambu-Goldstone bosons used as inflatons are volume moduli which enjoy a rescaling symmetry of the form:\cite{Burgess:2014tja}
\be
\Phi\to e^{\alpha}\,\Phi\qquad \Phi=\rho\,e^{i\theta}\qquad \rho\to e^{\alpha} \rho
\nonumber
\ee
The canonical normalisation $\rho = e^{\varphi/f}$ yields a non-periodic shift symmetry $\varphi\to\varphi+\alpha f$. Notice that the effective field theory is under control when $\rho\gg 1$ $\Leftrightarrow$ $\varphi\gg f$, implying that $\varphi\gg M_p$ is a natural regime for $f\sim M_p$. Moreover the decoupling of the fields orthogonal to the inflaton is easier because of the no-scale cancellation which gives a mass to the $S$ and $U$-moduli at tree-level keeping the $T$-moduli massless. The symmetry breaking effects which generate the inflaton potential look like $V_0 \,e^{\pm\varphi/f}$, yielding $V= V_0\left(1-e^{-\varphi/f}\right)$. The phenomenological implications of this kind of potentials are:\cite{Burgess:2016owb}
\be
\epsilon\simeq\frac 12 \left(\frac{f}{M_p}\right)^2\eta^2\quad\text{and}\quad \eta\simeq - \left(\frac{M_p}{f}\right)^2 e^{-\varphi/f}<0
\quad\Rightarrow\quad \epsilon\ll|\eta|\ll 1
\nonumber
\ee
\be
r\simeq 2 \left(\frac{f}{M_p}\right)^2 (n_s-1)^2 \quad\Rightarrow\quad r\simeq 0.003 \left(\frac{f}{M_p}\right)^2 
\quad\text{for}\quad n_s\simeq 0.96  
\nonumber
\ee
Let us list three models with a different $f$, and so different predictions for $r$:
\begin{itemize}
\item K\"ahler moduli inflation:\cite{Conlon:2005jm}  $f\sim M_p/\sqrt{\mathcal{V}}\ll M_p$ $\Rightarrow$ $r \sim 10^{-10}$ 
\item Fibre inflation:\cite{Cicoli:2008gp} $\qquad \qquad f\sim M_p$  $\Rightarrow$  $r \sim 0.005$
\item Poly-instanton inflation:\cite{Cicoli:2011ct} $f\sim M_p/\ln\mathcal{V}$ $\Rightarrow$   $r \sim 10^{-5}$
\end{itemize}

\subsection{Strings and power loss at large scales}

The typical potential of models where the inflaton is a K\"ahler modulus involves also a steepening region for large values of $\varphi$: 
$V=V_0\left( 1-e^{-\varphi/f} + \delta\,e^{+\varphi/f}\right)$. In particular the potential of Fibre inflation is very similar to the Starobinsky model since $f=\gamma f_{\rm Staro}$ with $\gamma=1/\sqrt{2}$. The corresponding dual version is $R^{2-\gamma}+R^2$. Hence Fibre inflation provides a scalar-tensor theory which is the prototype of a working UV completion of the Starobinsky model since $\delta\simeq g_s^4\ll 1$ explains why $R^{n>2}$-terms are suppressed.\cite{Burgess:2016owb}

Moreover the positive exponential term can provide an interesting explanation of a possible power loss at large angular scales.\cite{Cicoli:2013oba} In fact, after fitting Planck precision data at $\ell >50$, one can predict the CMB power at $\ell <50$, finding a suppressed power at low-$\ell$ with around a $10\%$ deficit at about $2 \sigma$. \cite{Ade:2015lrj} This can be obtained in string inflationary models if the positive exponential becomes important just after the first $60$ efoldings of inflation ($N_e\simeq g_s^{-4/3}\gg 1$ in Fibre inflation \cite{Cicoli:2008gp}). This gives a departure from slow-roll and subsequently, for larger values of $\varphi$, the effective field theory is not under control anymore. This steepening of the inflationary potential gives a power loss at low-$\ell$ which can be intuitively understood by looking at the slow-roll expression of the amplitude $A_S(\text{large scales})\simeq \frac{V^{3/2}}{M_p^3 V'}\ll 10^{-5}$. 

Interestingly, a power loss at large scales is a typical and generic feature of models of just enough inflation. \cite{Cicoli:2014bja}
In fact, a model-independent analysis of any non-slow-roll background evolution prior to slow-roll inflation has revealed a high degree of universality since a power loss at large scales occurs for most common backgrounds: fast-roll ($w=1$), matter ($w=0$) and radiation dominance ($w=1/3$). This loss of power is associated with a peak with oscillations around the start of inflation.

\section{Post-inflationary string cosmology}

\subsection{Reheating from moduli decay}

After canonical normalisation the potential for the moduli around the minimum can be written as $V=\frac 12 m^2 \phi^2$ with $m\sim m_{3/2}\sim M_{\rm soft}\sim \mathcal{O}(1)$ TeV. During inflation this potential receives an extra contribution of the form:
\be
V= \frac 12 m^2 \phi^2 + c H_{\rm inf}^2 (\phi-\phi_0)^2 \sim c H_{\rm inf}^2 (\phi-\phi_0)^2\quad\text{for}\quad m \ll H_{\rm inf}
\nonumber
\ee
Thus $\phi$ is displaced from $\phi = 0$ during inflation. The equation of motion $\ddot{\phi}+ 3 H \dot{\phi} + m^2\phi=0$ shows that $\phi$ behaves as a harmonic oscillator with friction. At the end of inflation the friction wins, and so $\phi$ is frozen at $\phi=\phi_0$. Reheating from the inflaton decay ($\phi$ is the lightest modulus different from the inflaton) leads to a thermal bath with temperature $T$ and $H\sim T^2/M_p$. The Universe expands and cools down, and so $H$ decreases. The field $\phi$ starts oscillating when $H \sim m$ and stores an energy of the order $\rho_\phi \sim m^2 \phi_0^2 \sim H^2 M_p^2 \sim T^4 \sim \rho_{\rm rad}$. However $\phi$ redshifts as matter as $\rho_\phi\propto T^3$ while the thermal bath redshifts as radiation as $\rho_{\rm rad}\propto T^4$. Thus $\phi$ quickly comes to dominate the energy density of the Universe, and so dilutes everything when it decays at $H\sim \Gamma \sim m^3/M_p^2$ giving rise to a reheating temperature of the order $T_{\rm rh}\sim\sqrt{\Gamma M_p}\sim m\sqrt{m/M_p}$. 
 
This picture leads to a non-standard cosmology from strings. Focussing on $m_\phi > 50$ TeV to have $T_{\rm rh} > T_{\rm BBN} \sim  3$ MeV, the decay of $\phi$ causes several modifications:
\begin{itemize}
\item Axionic dark matter is diluted if $T_{\rm rh} < \Lambda_{\rm QCD}\simeq 200$ MeV. If $T_{\rm rh} \gtrsim T_{\rm BBN}$ one can have $f_a \sim 10^{14}$ GeV without tuning the initial axion misalignment angle.\cite{Fox:2004kb}

\item Standard thermal LSP dark matter gets diluted if the reheating temperature is below the freeze-out temperature, i.e. $T_{\rm rh} < T_{\rm f} \simeq m_{\rm DM}/20 \sim \mathcal{O}(10)$ GeV.\cite{Allahverdi:2013noa, Acharya:2008bk}

\item Baryon asymmetry produced before $\phi$ decay also gets diluted. This can be a promising effect for Affleck-Dine baryogenesis which tends to be too efficient.\cite{Kane:2011ih}

\item Non-thermal dark matter gets produced from $\phi$ decay in two different ways:\cite{Allahverdi:2013noa}
\begin{enumerate}
\item \emph{Annihilation scenario} for $T_{\rm rh}$ close to $T_{\rm f}$: an abundant initial production of dark matter is followed by an efficient annihilation. In this case the LSP has to be Wino- or Higgsino-like.

\item \emph{Branching scenario} for $T_{\rm rh}$ close to $T_{\rm BBN}$: a smaller initial production of dark matter is followed by an inefficient annihilation. In this case the LSP has to be Bino-like.
\end{enumerate}
\end{itemize}

\subsection{Non-thermal dark matter}

In order to understand if the lightest modulus decay leads to thermal or non-thermal dark matter, one has to ask what is the generic value of $T_{\rm rh}$ from strings. This question can be answered by considering generic features of string compactifications:\cite{Allahverdi:2013noa, Acharya:2008bk}
\begin{enumerate}
\item SUSY breaking generates $m_\phi$;
\item Moduli mediate SUSY breaking to the MSSM via gravitational interactions, and so $M_{\rm soft}  = k\, m_\phi$ with $k$ a model-dependent constant of proportionality;
\item Since $m_\phi > 50$ TeV, one can get TeV-scale SUSY only for $k \ll 1$;  
\item In particular models one can have either $k \sim \mathcal{O}(10^{-2})$ from loop suppression factors or $k \sim \mathcal{O}(10^{-3} - 10^{-4})$ from sequestering effects;\cite{Aparicio:2014wxa}
\item For $M_{\rm soft} \sim \mathcal{O}(1)$ TeV, the reheating temperature can be written as:
\be
T_{\rm rh}\simeq m\sqrt{m/M_p} \sim k^{-3/2} M_{\rm soft} \sqrt{M_{\rm soft}/M_p} \sim k^{-3/2} \mathcal{O}(10^{-2})\, \text{MeV}
\nonumber
\ee
For $10^{-4}\leq k \leq 10^{-2}$ this gives $10 \,{\rm MeV}\leq T_{\rm rh} \leq 10$ GeV which is below the freeze-out temperature for LSP masses between $\mathcal{O}(100)$ GeV and $\mathcal{O}(1)$ TeV since:
\be
10\,{\rm GeV}\leq T_{\rm f} \simeq \frac{m_{\rm DM}}{20}\leq 100\,{\rm GeV}
\nonumber
\ee
\end{enumerate}                                                                        
Hence we conclude that string compactifications tend to give rise to non-thermal dark matter. Let us have a look at its production mechanism in the annihilation scenario. The decay of $\phi$ dilutes thermal dark matter enlarging the underlying parameter space, and reproduces dark matter non-thermally as follows:\cite{Allahverdi:2013noa}
\be
\frac{n_{\rm DM}}{s} = \left(\frac{n_{\rm DM}}{s}\right)_{\rm obs} \frac{\langle \sigma_{\rm ann} v\rangle^{\rm th}_{\rm f}}{\langle \sigma_{\rm ann} v\rangle_{\rm f}} \left(\frac{T_{\rm f}}{T_{\rm rh}}\right)
\nonumber
\ee
where $\left(\frac{n_{\rm DM}}{s}\right)_{\rm obs}\simeq 5\cdot 10^{-10} \left(\frac{1\,{\rm GeV}}{m_{\rm DM}}\right)$ and $\langle \sigma_{\rm ann} v\rangle^{\rm th}_{\rm f}\simeq 3\cdot 10^{-26}\,{\rm cm}^3\,{\rm s}^{-1}$. Clearly, in order to reproduce the observed value one needs $\langle \sigma_{\rm ann} v\rangle_{\rm f} = \langle \sigma_{\rm ann} v\rangle^{\rm th}_{\rm f} (T_{\rm f}/T_{\rm rh})$. Since $T_{\rm rh}<T_{\rm f}$, we have to consider the case with $\langle \sigma_{\rm ann} v\rangle_{\rm f} > \langle \sigma_{\rm ann} v\rangle^{\rm th}_{\rm f}$ leading to Wino/Higgsino-like LSP dark matter. For Bino-like LSP we have $\langle \sigma_{\rm ann} v\rangle_{\rm f} < \langle \sigma_{\rm ann} v\rangle^{\rm th}_{\rm f}$ which yields dark matter overproduction.

\subsection*{Non-thermal CMSSM}

Let us now study the phenomenological consequences of a non-standard cosmological history in the CMSSM case with non-thermal dark matter.\cite{Aparicio:2015sda} After imposing: 
\begin{enumerate}
\item Radiative EW symmetry breaking and a Higgs mass around 125 GeV;
\item No dark matter overproduction;
\item Present bounds from colliders (LHC), CMB (Planck), direct (LUX) and indirect (Fermi) dark matter searches; 
\end{enumerate}
the observed dark matter content turns out to be saturated for $T_{\rm rh}= 2$ GeV and a $300$ GeV Higgsino-like LSP.\cite{Aparicio:2015sda} Moreover the masses of the supersymmetric particles resemble a typical natural SUSY spectrum: $m_{\tilde{g}} \sim 2-3$ TeV, $m_{\tilde{t}}\sim 4-5$ TeV and the neutralinos $\tilde{\chi}_1^0$, $\tilde{\chi}_2^0$ and the chargino $\tilde{\chi}_1^+$ are almost degenerate. This model has a clear LHC signature: neutralino production via vector boson fusion.\cite{Dutta:2012xe} All this can be realised in string models with sequestered SUSY breaking.\cite{Aparicio:2014wxa}

\subsection{Axionic dark radiation}

A generic feature of string compactifications is the presence of light axionic degrees of freedom which is unavoidable in string models where not all the moduli are fixed by non-perturbative effect. \cite{Allahverdi:2014ppa} This leads to the production of axionic dark radiation from the decay of the lightest modulus. \cite{Cicoli:2012aq} In fact, the moduli are gauge singlets, and so they do not prefer to decay into visible sector fields and might have non-negligible branching ratios into light axions. This results in a non-zero contribution to the effective number of neutrino-like species $N_{\rm eff}$ which parameterises the energy density of radiation as:
\be
\rho_{\rm rad} = \rho_{\gamma} \left( 1 + \frac{7}{8}\left( \frac{4}{11} \right)^{4/3} N_{\rm eff} \right) 
\nonumber
\ee
However there are tight bounds from observations, $N_{\rm eff}=3.52^{+0.48}_{-0.45}$ at $95\%$ ${\rm CL}$,\cite{Ade:2013zuv} which would give a central value of order $\Delta N_{\rm eff}\simeq 0.5$. Planck 2015 data reduced the inferred amount of dark radiation to $N_{\rm eff} = 3.13 \pm 0.32$ at $68\% {\rm CL}$.\cite{Ade:2015xua} However one should take into account that Planck 2015 data are in slight tension with the HST value of $H_0$ together with the fact that $N_{\rm eff}$ is positively correlated with the value of the Hubble constant.

Thus when $\phi$ decays, it produces both SM particles and axionic dark radiation. These axions are relativistic, and so behave as radiation even if they are not in thermal equilibrium with SM particles since they are very weakly (gravitationally) coupled. Hence they free-stream to present day. Given that the temperature of the thermal bath is $T_\gamma \sim T_{\rm rh}\sim m_\phi\sqrt{m_\phi/M_p}$ while the energy of the axions is $E_a = m_\phi/2$, the ratio between the two energies is:\cite{Conlon:2013txa}
\be
\frac{E_a}{T_\gamma}\sim \sqrt{\frac{M_p}{m_\phi}}\sim 10^6\left(\frac{10^6\,{\rm GeV}}{m_\phi}\right)^{1/2}
\nonumber
\ee
This ratio is retained through all cosmic history. Therefore, if the lightest modulus mass is around $10^6$ GeV (often associated with low-energy SUSY in many string models), for $T_\gamma \sim 10^{-4}$ eV these axions today have an energy of order $100$ eV. Hence we have the prediction of a Cosmic Axion Background (CAB) with energies in the soft X-ray wavebands.\cite{Conlon:2013txa} Let us stress that this prediction comes from very general properties of string moduli since it relies just on the existence of massive particles with only gravitational couplings to ordinary matter. 

This CAB can be revealed via axion-photon conversion in coherent magnetic fields induced by a Lagrangian of the form:
\be
\mathcal{L} = -\frac 14 F^{\mu\nu}F_{\mu\nu} +\frac 12 \partial_\mu a \partial^\mu a -\frac 12 m_a^2 a^2 -\frac{a}{4 M} F^{\mu\nu}\tilde{F}_{\mu\nu}
\nonumber
\ee
Notice that $M \geq 10^{11}$ GeV from supernovae cooling bounds. The axion-photon conversion probability in a plasma with frequency $\omega_{\rm pl}$ is given by ($L$ is the coherence length of the magnetic field):\cite{Conlon:2013txa}
\begin{enumerate}
\item  $P_{a\to \gamma}\sim \frac 14 \left(\frac{B\,L}{M}\right)^2$ for $m_a < \omega_{\rm pl}$

\item $P'_{a\to \gamma}\sim P_{a\to \gamma}\left(\frac{\omega_{\rm pl}}{m_a}\right)^4\ll P_{a\to \gamma}$  for $m_a \gg \omega_{\rm pl}$  
\end{enumerate}
In order to have a large conversion probability we need therefore large values of $B$ and $L$. Promising astrophysical objects where this condition is satisfied are galaxy clusters which have a typical size of order $R_{\rm cluster} \sim 1$ Mpc and $B \sim 1 - 10$ $\mu G$ with $L \sim 1 - 10$ kpc. The ICM plasma frequency is of order $\omega_{\rm pl} \sim 10^{-12}$ eV, implying that axions with $m_a \gg 10^{-12}$ eV (like the QCD axion) give rise to a negligible conversion probability. 
    
\subsection*{CAB evidence in the sky}

A substantial soft X-ray excess in galaxy clusters above the thermal emission from the ICM has been observed since 1996 by several missions (EUVE, ROSAT, XMM-Newton, Suzaku and Chandra).\cite{Durret:2008jn} Its statistical significance is very large and at present there is no astrophysical explanation which is completely satisfactory. The typical excess luminosity is about $\mathcal{L}_{\rm excess}\sim 10^{43}$ erg s$^{-1}$. Given that the CAB energy density is $\rho_{\rm CAB}=1.6\cdot 10^{60}$ erg Mpc$^{-3}\left(\frac{\Delta N_{\rm eff}}{0.57}\right)$, the soft X-ray luminosity from axion-photon conversion becomes:
\be
\mathcal{L}_{a\to \gamma} = \rho_{\rm CAB}\,P_{a\to\gamma}^{\rm cluster} = 3.16\cdot 10^{43}\,{\rm erg}\,{\rm s}^{-1}\left(\frac{\Delta N_{\rm eff}}{0.5}\right)\left(\frac{B}{\sqrt{2}\mu G}\frac{10^{12}\,{\rm GeV}}{M}\right)^2\left(\frac{L}{1\,{\rm kpc}}\right)
\nonumber
\ee
This can match the data for $\Delta N_{\rm eff}\simeq 0.5$, $m_a<10^{-12}$ eV and $M\sim 10^{12}$ GeV.\cite{Conlon:2013txa}

\subsection*{The 3.5 keV line}

Recently several missions have claimed the detection of a $3.5$ keV line from:
\begin{enumerate}
\item Stacked galaxy clusters (XMM-Newton) and Perseus (Chandra);\cite{Bulbul:2014sua}
\item Perseus and Andromeda (XMM-Newton);\cite{Boyarsky:2014jta}
\item Perseus (Suzaku);\cite{Urban:2014yda}
\end{enumerate}
On the contrary, this $3.5$ keV line has not been detected from:
\begin{enumerate}
\item Dwarf spheroidal galaxies (XMM-Newton);\cite{Malyshev:2014xqa}
\item Stacked galaxies (XMM-Newton and Chandra);\cite{Anderson:2014tza}
\end{enumerate}
The origin of this line could be astrophysical if it is due to a new atomic transition in the ICM plasma. On the other hand, the simplest particle physics interpretation involves a dark matter particle with mass $m_{\rm DM}\sim 7$ keV (main candidates are sterile neutrinos, axions and axinos) decaying into photons. However there are a few problems with this simple explanation:
\begin{enumerate}
\item Inconsistent inferred signal strength: the line should trace only the dark matter quantity in each cluster yielding a clear prediction:
\be
F_{{\rm DM}\to \gamma}^i\propto \Gamma_{{\rm DM}\to \gamma} \rho_{\rm DM}^i\quad\Rightarrow\quad   
\frac{F_{{\rm DM}\to \gamma}^i}{F_{{\rm DM}\to \gamma}^j} \propto \frac{\rho_{\rm DM}^i}{\rho_{\rm DM}^j}\quad\text{fixed}
\nonumber
\ee
Nonetheless the signal strength from Perseus is larger than for other galaxy clusters,\cite{Bulbul:2014sua,Boyarsky:2014jta} and Coma, Virgo and Ophiuchus.\cite{Urban:2014yda}

\item  Inconsistent morphology of the signal: one would expect a non-zero signal from everywhere in the dark matter halo but the signal is stronger from the central cool core of Perseus,\cite{Bulbul:2014sua,Boyarsky:2014jta,Urban:2014yda} and Ophiucus and Centaurus.\cite{Boyarsky:2014jta}

\item Non-observation in dwarf spheroidal galaxies: dwarf galaxies are dominated by dark matter, and so they should give the cleanest dark matter decay line but no line has been observed from these astrophysical objects.
\end{enumerate}

\subsection*{Alternative explanation: DM $\to$ ALP $\to \gamma$}

An alternative explanation of the $3.5$ keV line relies on a monochromatic $3.5$ keV axion line produced from the decay of a dark matter particle with $m_{\rm DM} \sim 7$ keV, followed by axion-photon conversion in the cluster magnetic field.\cite{Cicoli:2014bfa} The dark matter decay into axions could be induced by couplings of the form:
\be
\text{a)} \quad \frac{\Phi}{\Lambda}\partial_\mu a \partial^\mu a\qquad\qquad\Rightarrow\qquad\qquad \Gamma_\Phi = \frac{1}{32\pi}\frac{m_\Phi^3}{\Lambda^2}
\nonumber
\ee
\be
\text{b)}\quad \frac{\partial_\mu a}{\Lambda} \bar{\psi}\gamma^\mu \gamma^5\chi \quad\Rightarrow\quad 
\Gamma_{\psi\to\chi a}= \frac{1}{16\pi}\frac{(m_\psi^2-m_\chi^2)^3}{m_\psi^3\Lambda^2}
\nonumber
\ee
The predicted photon flux is:\cite{Cicoli:2014bfa}
\be
F_{{\rm DM}\to \gamma}^i\propto \Gamma_{{\rm DM}\to a} P_{a\to \gamma}^i \rho_{\rm DM}^i\quad\Rightarrow\quad   
\frac{F_{{\rm DM}\to \gamma}^i}{F_{{\rm DM}\to \gamma}^j} \propto \frac{\rho_{\rm DM}^i P_{a\to \gamma}^i}{\rho_{\rm DM}^j P_{a\to \gamma}^j}
\propto\left(\frac{B^i}{B^j}\right)^2
\nonumber
\ee
It is interesting to notice that observational data can be matched for the same values which reproduce the soft X-ray excess: $m_a<10^{-12}$ eV and $M\sim 10^{12}$ GeV. Moreover this line with a $B$-dependent strength can explain all the anomalies listed above:
\begin{enumerate}
\item Since the photon flux depends on both dark matter density and $B$-field, a stronger signal from Perseus is explained by its large magnetic field
\item The morphology of the signal is explained by the fact that the $B$-field peakes at the central cool core in galaxy clusters
\item The non-observation from dwarf galaxies is due to the fact that $L$ and $B$ are smaller than in galaxy clusters. This has been predicted in Ref.\refcite{Cicoli:2014bfa} and afterwards confirmed in Ref. \refcite{Malyshev:2014xqa}
\item The non-observation in galaxies is again due to the fact that $L$ and $B$ are smaller than in galaxy clusters. This effects has also been predicted in Ref.\refcite{Cicoli:2014bfa} and afterwards confirmed in Ref. \refcite{Anderson:2014tza}
\item The observation of the line in Andromeda could be due to the fact that this galaxy is almost edge on to us, and so axions have significant passage through its disk enhancing their conversion probability before reaching us.
\end{enumerate}

\section{Conclusions}

Let us list the main topics discussed in this talk:
\begin{itemize}
\item Globally consistent chiral models with full closed string moduli stabilisation;
\item dS vacua compatible with the presence of chiral matter;
\item Phenomenological applications: SUSY breaking, TeV-scale soft terms, inflation, dark matter and dark radiation   
\item Difficulty to build robust inflationary models with detectable tensor modes since this requires $\Delta\phi > M_p$
\item K\"ahler moduli as promising inflaton candidates due to the presence of an effective shift symmetry from the extended no-scale structure
\item Largest value of tensor-to-scalar ratio of order $r \leq 0.01$ in models where the inflaton is a K\"ahler modulus
\item Generic power loss at large scales for models of just enough inflation
\item Reheating driven by the decay of the lightest modulus
\item Non-standard cosmology characterised by the dilution of thermal dark matter
\item Production of non-thermal dark matter 
\item Non-thermal CMSSM with a $300$ GeV Higgsino-like LSP saturating the dark matter content for $T_{\rm rh} = 2$ GeV
\item Generic production of axionic dark radiation
\item Prediction of a cosmic axion background with $E_a \sim 200$ eV that is detectable via axion-photon conversion in astrophysical magnetic fields
\item The soft X-ray excess and  the $3.5$ keV line from galaxy clusters could be due to axion-photon conversion.

\end{itemize}

\section*{Acknowledgments}

I would like to thank R. Allahverdi, L. Aparicio, C. Burgess, J. Conlon, B. Dutta, K. Dutta, S. Downes, D. Klevers, S. Krippendorf, A. Maharana, C. Mayrhofer, D. Marsh, F. Muia, F. Pedro, F. Quevedo, M. Rummel, K. Sinha, R. Valandro, A. Westphal and M. Williams for fruitful collaboration on the topics covered in this talk.

\end{document}